\documentclass[a4paper,fleqn,10pt]{article}
\usepackage{graphicx}
\usepackage{subfigure}
\usepackage{latexsym}
\usepackage{amsmath,amssymb}
\usepackage{dcolumn}

\begin{document}

\title{Modified holographic Ricci dark energy coupled to interacting relativistic and non-relativistic dark matter in the nonflat universe}
\author {En-Kun Li,$^a$
Yu Zhang,$^{a,}$\thanks{Email: zhangyu\_128@126.com}
Jin-Ling Geng $^a$\\
$^a$ Faculty of Science, Kunming University of Science and Technology,\\Kunming 650500,China.}

\date{}
\maketitle

\begin{abstract}
The modified holographic Ricci dark energy coupled to interacting relativistic and non-relativistic dark matter is considered in the nonflat Friedmann-Robertson-Walker universe.
Through examining the deceleration parameter, one can find that the transition time of the Universe from decelerating to accelerating phase in the interacting holographic Ricci dark energy model is close to that in the $\Lambda$ cold dark matter model.
The evolution of modified holographic Ricci dark energy's state parameter and the evolution of dark matter and dark energy's densities shows that the dark energy holds the dominant position from the near past to the future.
By studying the statefinder diagnostic and the evolution of the total pressure, one can find that this model could explain the Universe's transition from the radiation to accelerating expansion stage through the dust stage.
According to the $Om$ diagnostic, it is easy to find that when the interaction is weak and the proportion of relativistic dark matter in total dark matter is small, this model is phantom-like.
Through our studying, we find the interaction and the relativistic dark matter's proportion all have great influence on the evolution of the Universe.
\end{abstract}


\section{Introduction}
\label{sin}

It is widely accepted that our Universe is undergoing an accelerating expansion. Observations from type Ia supernovae \cite{Riess1998,Perlmutter1999}, cosmic microwave background radiation \cite{Spergel2003,Spergel2007}, and Sloan Digital Sky Survey \cite{Tegmark2004,Seljak2005}, have given supports to the present accelerating cosmic expansion. Data from these observations have suggested that nearly $25\%$ of the total matter-energy in the Universe is referred to as dark matter (DM), and more than $70\%$ of the total matter energy is referred to as dark energy (DE) \cite{Bradac2008,Caldwell2009,Li2011}. The mysterious DM is helpful in explaining the galactic curves and structure formation in the Universe \cite{Zhang2011}.
A common assumption of DM is generically referred to as cold dark matter (CDM) or non-relativistic DM, which moves at a so small speed compared to the light speed that it has no relativistic effects \cite{Peebles1982,Blumenthal1984}.
Although the $\Lambda$CDM model of cosmology is found to be an indisputable success on large scale, it still has several significant indications of possible shortcomings on a smaller scale \cite{Cole2005,Percival2010,Komatsu2011,Mahdi2014}.
Nowadays, observations show that the CDM galaxy haloes contain many more satellites around the Milky Way and M$31$ than the observed satellites \cite{Moore1999,Klypin1999}; the inner density profiles used to fit simulated CDM halos are denser than that inferred from the rotation curves of real galaxies \cite{Simon2005,Gentile2009}; the expected number of galaxies in CDM is bigger than the observed \cite{Peebles2010,Anderhalden2010}.
By now, a lot of efforts have been dedicated to solve these problems \cite{Maddox1990,Davis1992,Diemand2008,Springel2008,Stadel2009}, and the type of non-relativistic DM mixed with relativistic DM has attracted a lot of attention \cite{Lovell2011,Polisensky2011,Colin2008,Maccio2012}.
Some observations also suggested that the total DM does not only have a non-relativistic component but also has a relativistic component \cite{Wyman2014,Hamann2013,Battye2014,Gariazzo2013}. Then, in the present paper, we would like to take the mixed dark matter into consideration.

DE is another important component of the Universe's total matter energy, whose pressure is negative, pushing the Universe into accelerating expansion \cite{Kamensshchik2001,Dev2003,Alcaniz2004,Brevik2004}. As is well known, the cosmological constant is the simplest explanation for the DE phenomenon \cite{Weinberg1989,Sahni2000,Peebles2003,Padmanabhan2003,Nobbenhuis2006}. However, the ``fine-tuning problem'' and the ``cosmic coincidence'' problems arise from the cosmological constant scenario \cite{Copeland2006}. In order to solve these two problems, so many other DE models have been put forward, such as quintessence \cite{Wetterich1988,Steinhardt1999,Sahni2002}, phantom \cite{Caldwell2002,Nojiri2003,Feng2010}, quintom \cite{Elizalde2004,Feng2005,Guo2005}, Chaplygin gas \cite{Wu2007,Sadjadi2010,Xu2010a}, and so forth. More detailed information about the DE models can be found in the works of \cite{Padmanabhan2005,Bamba2012}. Apart from these models, another popular DE model referred to as ``holographic dark energy,'' which arises from the holographic principle \cite{Hooft}, was proposed in works like \cite{Horava2000,Thomas2002,Hsu2004}. Holographic dark energy models provide a more simple and reasonable frame to investigate the problem of DE \cite{Setare2006,Sheykhi2009}. The energy density of holographic dark energy is given by $\rho_{de}=3c^2M^2_{pl}L^{-2}$, where $L$ indicates the infrared (IR) cutoff radius, $M_{pl}=1/\sqrt{8\pi G}$ is the Planck mass, and $c$ is a numerical constant \cite{Li2004}. The IR cutoff has been considered as the Hubble radius \cite{Cohen1999,Horava2000,Thomas2002}, the particle horizon \cite{Fischler,Bousso1999}, the future event horizon \cite{Li2004,Huang2004}, the cosmological conformal time \cite{Cai2007,Wei2008}, or other generalized IR cutoff \cite{Elizalde2005,Gao2009,Granda2008,Granda2009,Chen2009,Cai2009,Xu2009,Duran2011,Chimento2011}. Among them, Gao $\it{et}$ $\it{al.}$ \cite{Gao2009} raised a holographic Ricci DE model, whose length scale is the inverse of the Ricci curvature scalar, i.e., $L\thicksim|R|^{-1/2}$. Granda and Oliveros \cite{Granda2008,Granda2009} suggested a new holographic Ricci DE model with the density of DE as $\rho_{de}=3M^2_{pl}(\alpha H^2+\beta\dot{H})$. In the same year, Chen and Jing \cite{Chen2009} modified this model as $\rho_{de}=3M^2_{pl}(\alpha H^2+ \beta \dot{H} +\gamma \ddot{H} H^{-1})$. Moreover, the holographic dark energy models have been tested by various cosmic observations \cite{Xu2009a,Xu2010,Zhang2009,Micheletti2010,Wang2010}.

Since DM and DE are both the mysterious elements in the Universe, the physics behind the dark sector has become an interesting research field in modern cosmology, and the models with the interaction between DM and DE have gained great attention \cite{Wei2007,Quartin2008,Karwan2008,Chimento2010,Chimento2012,Zhang2010}. Especially, at the present time, there are many works that have been done in discriminating the behaviors of different holographic dark energy models and interacting holographic dark energy models \cite{Zhang2005,Setare2007,Zhang2008,Feng2008}. Recently, Chimento $\it{et}$ $\it{al.}$ have done some excellent works with regard to the modified holographic Ricci dark energy (MHRDE), and they find that the MHRDE coupled to the interacting DM can induce a relaxed Chaplygin gas \cite{Chimento2011}. Using the $\chi^2$ method to the observational Hubble data, Chimento $\it{et}$ $\it{al.}$ find it is consistent with the bound $\Omega_{de} (z\simeq1100)< 0.1$ reported for the behavior of DE at an early stage \cite{Chimento2012}, and more of their work about the interacting modified holographic Ricci dark energy (IMHRDE) model can be found in \cite{Chimento2013A,Chimento2013B}. The MHRDE interacting with pressureless non-relativistic DM in a flat Friedmann-Robertson-Walker (FRW) universe has been studied by Chattopadhyay $\it{et}$ $\it{al.}$ \cite{Chattopadhyay2013}, and they found that this model is able to attain the $\Lambda$CDM phase of the Universe. In addition, current observations from type Ia supernovae and cosmic microwave background radiation, etc., also support the proposal of a possible interaction between DM and DE \cite{Guo2007}. Moreover, some experimental data and recent papers have implied that our Universe is not a perfectly flat universe, but with spatial curvature \cite{Spergel2003,Tegmark2004,Bennett2003,Seljak2006,Spergel2007}. Then, in the present paper, we would like to investigate a nonflat universe composed of interacting relativistic and non-relativistic DM and MHRDE. The difference between the relativistic DM and the non-relativistic DM is whether there are obvious relativistic effects. The model of MHRDE considered in the present paper is introduced by Granda and Oliveros \cite{Granda2009}. The energy density of the MHRDE with an IR cut off is given by \cite{Granda2009,Chimento2011}
\begin{eqnarray}
&\rho_{de}=\frac{2}{\alpha-\beta}(\dot{H}+\frac{3}{2}\alpha H^2),\label{ede}
\end{eqnarray}
where $\alpha$ and $\beta$ are two constants, $H=\dot{a}/a$ is the Hubble parameter, and $\cdot$ represents $d/dt$. In the present paper we choose $8\pi G=c=1$.

The present article is outlined as follows. In Sec. \ref{seq}, we give the basic equations and solutions for the IMHRDE model. In Sec. \ref{sev}, we would like to examine the evolution of the Universe with the IMHRDE model. We will study the evolution of the equation of state (EOS) parameter for MHRDE and deceleration parameter. The diagnostic of statefinder parameters $\{r,s\}$ \cite{Sahni2003,Evans2005,Setare2007} and $Om$ parameter \cite{Sahni2008} of the IMHRDE model are also studied. In Sec. \ref{scon}, we give our conclusions.

\section{Basic equations and solutions of the IMHRDE model}
\label{seq}

The line element of the nonflat FRW universe is given by
\begin{eqnarray}
  &ds^2=-dt^2+a^2(t)\left[\frac{dr^2}{1-kr^2}+r^2(dr^2+ \sin^2\theta d\varphi^2)\right],
  \label{eF}
\end{eqnarray}
where $a(t)$ represents the dimensionless scale factor, $k$ denotes the curvature of space and $k=0,1,-1$ corresponds to flat, closed and open FRW universe, respectively.

Now, consider the Universe filled with DM and MHRDE, and the total DM has two components: the pressure relativistic one and the pressureless non-relativistic one. In such conditions, the evolution of the Universe described by the Friedmann equation can be written as
\begin{eqnarray}
  &H^2+\frac{k}{a^2}=\frac{1}{3}(\rho_{dm}+\rho_{de})=\frac{1}{3} (\rho_{rdm}+\rho_{nrdm}+\rho_{de}),
  \label{eFr}
\end{eqnarray}
where $\rho_{dm}=\rho_{rdm}+\rho_{nrdm}$ and $\rho_{de}$ are the energy densities for total DM and MHRDE, $\rho_{rdm}$ is the density of relativistic DM and $\rho_{nrdm}$ is the density of non-relativistic DM. Here, in this paper, we assume that the ratio between the density of relativistic DM and non-relativistic DM is a fixed value. Suppose that $\rho_{rdm}=\gamma\rho_{dm}$, then $\rho_{nrdm}=(1-\gamma)\rho_{dm}$, where $\gamma$ is a constant. In order to preserve the local energy-momentum conservation law, i.e. $\nabla_{\mu}T^{\mu\nu}=0$, the following equation must be satisfied:
\begin{eqnarray}
  &\dot{\rho}_{tot}+3H(\rho_{tot}+p_{tot})=0,
  \label{ertot}
\end{eqnarray}
where $\rho_{tot}=\rho_{dm}+\rho_{de}$ and $p_{tot}=p_{dm}+p_{de}$ are  the total energy density and pressure, $p_{rdm}=w_{rdm} \rho_{rdm} =\gamma w_{rdm} \rho_{dm}=p_{dm}$ and $p_{de}=w_{de}\rho_{de}$ are the pressure of relativistic DM and MHRDE, and $w_{rdm}$ and $w_{de}$ are EOS parameters for relativistic DM and MHRDE, respectively. Considering Eq. (\ref{ertot}), the continuity equations of energy densities with $Q$ as the interaction are given by
\begin{eqnarray}
  &\dot{\rho}_{de}+3H(\rho_{de}+p_{de})=-Q,
  \label{edre}\\
  &\dot{\rho}_{dm}+3H(\rho_{dm}+w_{rdm}\rho_{rdm})=Q,
  \label{edrm}
\end{eqnarray}
where $Q$ denotes the interaction between DM and MHRDE. In the present paper, we take the interaction form as $Q=3bH(\rho_{de}+\rho_{dm}) =9bH (H^2+ \frac{k}{a^2})$ with the coupling constant $b$ \cite{Sheykhi2010}. For the total DM is constituted of relativistic DM and non-relativistic DM, Eq. (\ref{edrm}) could be written in the following two forms, respectively:
\begin{eqnarray}
  &\dot{\rho}_{rdm}+3H(1+\gamma w_{rdm})\rho_{rdm} =9\gamma b H (H^2+\frac{k}{a^2}),
  \label{edrrm}
\end{eqnarray}
or
\begin{eqnarray}
  &\dot{\rho}_{nrdm}+3H(1+\gamma w_{rdm})\rho_{nrdm} =9(1-\gamma)b H (H^2+\frac{k}{a^2}).
  \label{edrnm}
\end{eqnarray}

Here, we define that
\begin{eqnarray}
  &h=\frac{H}{H_0},\quad \tilde{\rho}_{de}=\frac{\rho_{de}}{3H_0^2}, \quad  \tilde{\rho}_{dm}=\frac{\rho_{dm}}{3H_0^2},
   \quad \tilde{\rho}_{rdm}=\frac{\rho_{rdm}}{3H_0^2}, \quad \tilde{\rho}_{nrdm} =\frac{\rho_{nrdm}}{3H_0^2},
  \label{etil}
\end{eqnarray}
where $H_0$ is the present value of the Hubble parameter. Meanwhile, we define that $\Omega_{k0}= k/H_0^2$, $\Omega_{dm0}= \rho_{dm0}/3H_0^2$ and $\Omega_{de0}= \rho_{de0}/3H_0^2$, which correspond to the present value of the fractional energy densities for curvature, DM and MHRDE, respectively. Then, according to Eq. (\ref{eFr}), we can obtain that $1+\Omega_{k0}=\Omega_{dm0}+\Omega_{de0}$.

Substituting Eq. (\ref{ede}) into Eq. (\ref{eFr}) and using Eq. (\ref{etil}), we obtain that
\begin{eqnarray}
  &h^2+\Omega_{k0}e^{-2x}=\tilde{\rho}_{dm}+\frac{\alpha} {\alpha-\beta}h^2 +\frac{1}{3(\alpha-\beta)}\frac{d h^2}{dx},
  \label{eth}
\end{eqnarray}
where $x=\ln a$ and $'$ represents $d/dx$. Differentiate this equation once more, one has $\tilde{\rho}'_{dm}$ as
\begin{eqnarray}
  &\tilde{\rho}'_{dm}=-\frac{\beta}{\alpha-\beta}\frac{dh^2} {dx}-\frac{1}{3(\alpha-\beta)}\frac{d^2h^2}{dx^2}-2\Omega_{k0}e^{-2x}.
  \label{etdmx}
\end{eqnarray}
Using the relation: $\dot{\rho}_{dm}=H\tilde{\rho}'_{dm}$, and Eqs. (\ref{edrm}), (\ref{eth}), and (\ref{etdmx}), we have
\begin{eqnarray}
  &\frac{d^2h^2}{dx^2} +3(1+\gamma w_{rdm} +\beta) \frac{dh^2} {dx}
  +9[(1+\gamma w_{rdm})\beta +b(\alpha-\beta)]h^2 \nonumber\\
   &=3(\alpha-\beta)(1+3\gamma w_{rdm}-3b)\Omega_{k0}e^{-2x}.
  \label{eq:d2h}
\end{eqnarray}
After some calculations, the general solution of the above differential equation can be written in the following form
\begin{eqnarray}
  &h^2=c_1e^{-\frac{3}{2}m_1x}+c_2e^{-\frac{3}{2}m_2x}+ce^{-2x},
  \label{eq:h2}
\end{eqnarray}
where
\begin{eqnarray}
  &m_{1,2}= (1+\gamma w_{rdm}+\beta) \pm\sqrt{(1 +\gamma w_{rdm}-\beta)^2-4b(\alpha-\beta)},
  \label{eq:m12}\\
  &c= -\Omega_{k0} +\frac{(3\alpha-2)(1+3\gamma w_{rdm})\Omega_{k0}}{(1+3\gamma w_{rdm})(3\beta-2)+9b(\alpha-\beta)}.
  \label{eq:c}
\end{eqnarray}
Define that $c=c_3-\Omega_{k0}$, then using Eq. (\ref{eq:c}) one can obtain that
$$c_3=\frac{(3\alpha-2)(1+3\gamma w_{rdm})\Omega_{k0}}{(1+3\gamma w_{rdm})(3\beta-2)+9b(\alpha-\beta)}.$$
Therefore, Eq. (\ref{eq:h2}) could be rewritten as
\begin{eqnarray}
&h^2+\Omega_{k0}e^{-2x}=c_1e^{-\frac{3}{2}m_1x}+c_2e^{-\frac{3}{2}m_2x}+c_3e^{-2x}.
\label{eq:h2x}
\end{eqnarray}
The coefficients $c_1$ and $c_2$ can be determined by the  following initial conditions:
\begin{eqnarray}
&h^2|_{x=0}=1,\quad\frac{dh^2}{dx}|_{x=0}=3(\alpha-\beta)\Omega_{de0}-3\alpha.
\label{eq:initial}
\end{eqnarray}
Then the coefficients can be written as
\begin{eqnarray}
  &c_1=\frac{6[(\alpha-\beta)\Omega_{de0}-\alpha]-(3m_2-4)(c_3-\Omega_{k0})+3m_2}{3(m_2-m_1)},\\
  \label{eq:c1}
  &c_2=-\frac{6[(\alpha-\beta)\Omega_{de0}-\alpha]-(3m_1-4)(c_3-\Omega_{k0})+3m_1}{3(m_2-m_1)}.
  \label{eq:c2}
\end{eqnarray}
Taking the relation $a=1/(1+z)$ into consideration, one can obtain that
\begin{eqnarray}
  &h^2=& c_1(1+z)^{\frac{3}{2}m_1}+c_2(1+z)^{\frac{3}{2}m_2}  +(c_3-\Omega_{k0})(1+z)^2,
  \label{eq:h2z}\\
  &\tilde{\rho}_{de}=& \frac{1}{\alpha-\beta} [(\alpha-\frac{m_1} {2})c_1(1+z)^{\frac{3}{2}m_1}  +(\alpha-\frac{m_2}{2})c_2(1+z)^{\frac{3}{2}m_2} \nonumber\\&&+(\alpha-\frac{2}{3}) (c_3-\Omega_{k0})(1+z)^2],
  \label{eq:rhoDz}\\
  &\tilde{\rho}_{dm}=& -\frac{1}{\alpha-\beta} [(\beta -\frac{m_1}{2})c_1(1+z)^{\frac{3}{2}m_1}  +(\beta-\frac{m_2}{2})c_2(1+z)^{\frac{3}{2}m_2} \nonumber\\&& +(\beta-\frac{2}{3})(c_3-\Omega_{k0})(1+z)^2]   +\Omega_{k0}(1+z)^2.
  \label{eq:rhoMz}
\end{eqnarray}
From these equations, it is easy to obtain the energy density of the relativistic or the non-relativistic DM with $\rho_{rdm}=\gamma\rho_{dm}$ and $\rho_{nrdm}=(1-\gamma)\rho_{dm}$.

Considering the basic equations and solutions  given above, we find that there are three free parameters, i.e., $b$, $\alpha$, and $\beta$, that should be determined. For this purpose, using Eqs. (\ref{eFr}), (\ref{edre}), and (\ref{edrm}), the derivative of $H$ with respect to time is given by
\begin{eqnarray}
&\dot{H}=\frac{k}{a^2}-\frac{3}{2}(H^2+\frac{k}{a^2}) \left(1+\frac{w_{de}+u\gamma w_{rdm}}{1+u}\right),
\label{eq:dotH}
\end{eqnarray}
where $u= \rho_{dm}/ \rho_{de}= \tilde{\rho}_{dm}/\tilde{\rho}_{de}$  is the ratio between the energy densities of DM and MHRDE. Substituting Eqs. (\ref{ede}) and (\ref{eq:dotH}) into Eq. (\ref{eFr}), one can easily find the relationship between $u$ and $w_{de}$ as
\begin{eqnarray}
&h^2= -\Omega_{k0}e^{-2x}+\frac{1+u}{3(\alpha-\beta)} [3\alpha h^2+2\Omega_{k0}e^{-2x} -3(h^2+\Omega_{k0}e^{-2x})(1+\frac{w_{de}+u\gamma w_{rdm}} {1+u})]. \label{eq:uw}
\end{eqnarray}
Now, taking the boundary conditions $w_{de0}=-1$ and $u_0=(1+\Omega_{k0}-\Omega_{de0})/\Omega_{de0}$ into consideration \cite{Yu2010}, and using Eq. (\ref{eq:uw}), we obtain
\begin{eqnarray}
&\beta=(1-\frac{1}{\Omega_{de0}})\alpha+(1+\gamma w_{rdm})u_0-\frac{2}{3}\frac{\Omega_{k0}}{\Omega_{de0}},
\label{eq:alphabeta}
\end{eqnarray}
the value of $\beta$ is given in terms of the free parameter $\alpha$, which has nothing to do with the coupling constant $b$. From Eq. (\ref{eq:alphabeta}), we find the coefficient before $\alpha$ is $(1-\frac{1}{\Omega_{de0}})$, and because $\Omega_{de0} <1$, we get that $\beta$ increases as $\alpha$ decreases. Now, the free parameters have been reduced to two, and they will be fixed by the behavior of the deceleration parameter, in particular, when $\alpha=4/3$, $\rho_{de} \propto R$, where $R=6(\dot{H}+2H^2)$ is the Ricci scalar curvature for a spatially flat FRW space-time.

\section{Evolution of the Universe with MHRDE and DM}
\label{sev}

\subsection{EOS and deceleration parameter}
\label{sEd}

In this section, we would like to examine the Universe's evolution by studying the evolution of the EOS parameter of MHRDE and the deceleration parameter. Using Eq. (\ref{edre}), one can obtain the fractional pressure of the MHRDE: $$\tilde{p}_{de}=\frac{p_{de}}{3H_0^2}=-\tilde{\rho}_{de}-\frac{1}{3}\frac{d\tilde{\rho}_{de}} {dx}-b(h^2+\Omega_{k0}e^{-2x}).$$
Then, the EOS parameter of MHRDE would be
\begin{eqnarray}
&w_{de}=\frac{p_{de}}{\rho_{de}}= \frac{\tilde{p}_{de}}{\tilde{\rho}_{de}} =-1-\frac{1}{3}\frac{d\ln\tilde{\rho}_{de}}{dx} -\frac{b(h^2+\Omega_{k0}e^{-2x})}{\tilde{\rho}_{de}}.
\label{eq:wD}
\end{eqnarray}
Substituting Eqs. (\ref{eq:h2z}) and (\ref{eq:rhoDz}) into Eq. (\ref{eq:wD}), we obtain
\begin{eqnarray}
  &w_{de}=-1-\frac{\gamma_1}{2\gamma_2}-\frac{2b\gamma_3}{\gamma_2},\label{eq:wDz}
\end{eqnarray}
where
\begin{eqnarray}
  &\gamma_1= -(2\alpha-m_1)m_1c_1(1+z)^{\frac{3}{2}m_1} -(2\alpha-m_2)m_2c_2(1+z)^{\frac{3}{2}m_2} -\frac{8}{3}(\alpha-\frac{2}{3})(c_3-\Omega_{k0})(1+z)^2,\\
  &\gamma_2= (2\alpha-m_1)c_1(1+z)^{\frac{3}{2}m_1} +(2\alpha-m_2)c_2(1+z)^{\frac{3}{2}m_2} -2(\alpha-\frac{2}{3})(c_3-\Omega_{k0})(1+z)^2,\\
  &\gamma_3= (\alpha-\beta)[c_1(1+z)^{\frac{3}{2}m_1}   +c_2(1+z)^{\frac{3}{2}m_2}+c_3(1+z)^2].
\end{eqnarray}
\begin{figure*}
\centering
   \includegraphics[scale=.59]{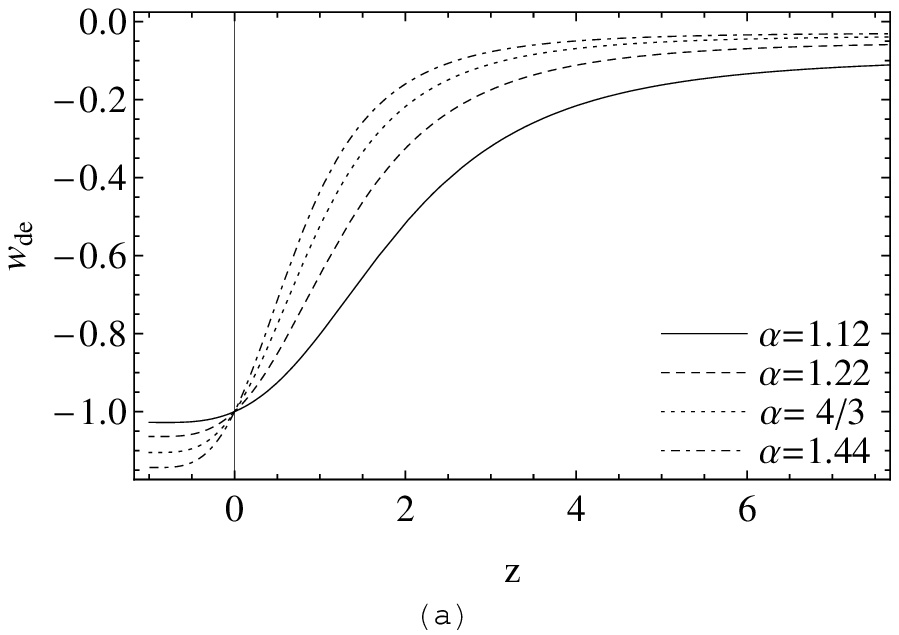}
   \includegraphics[scale=.59]{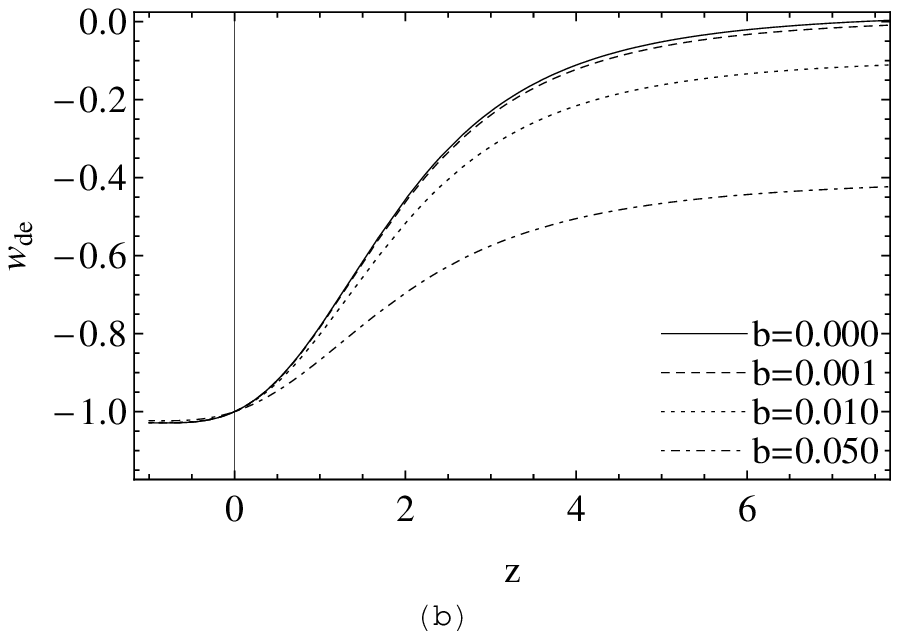}
   \includegraphics[scale=.59]{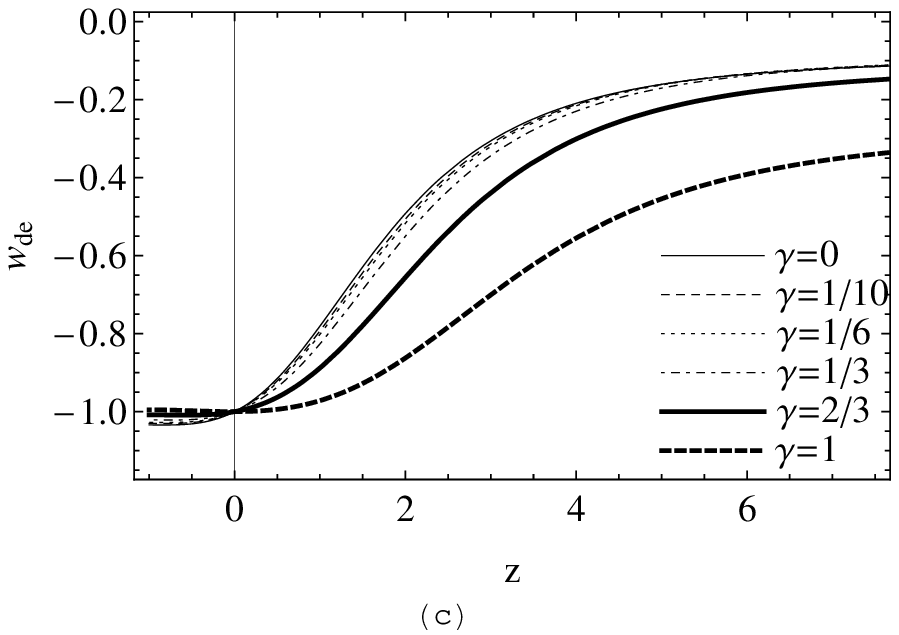}
\caption{The evolution of the EOS parameter of MHRDE, i.e., $w_{de}$, with respect to the redshift $z$ under different cases: (a) $b=0.01$, $\gamma=1/6$; (b) $\alpha=1.12$, $\gamma=1/6$; (c) $\alpha=1.12$, $b=0.01$. Here, we choose $\Omega_{k0}=0.02$, $w_{rdm}=0.1$, and $\Omega_{de}=0.73$.}
\label{fa}
\end{figure*}
We have plotted the evolution of the EOS parameter of MHRDE with respect to the redshift $z$ in Fig. \ref{fa}. Figure \ref{fa}(a) is for different $\alpha$ with fixed coupling constant $b= 0.01$ and fixed proportion parameter $\gamma= 1/6$. This figure shows that the value of $w_{de}$ approaches $0$ at high redshift, which indicates that the MHRDE behaves like DM in the past stage of the Universe. At low redshift or in the future, the values of $w_{de}$ depend on $\alpha$, for bigger $\alpha$, the value of $w_{de}$ is smaller. In Fig. \ref{fa}(a), one can easily find that for different $\alpha$, the value of $w_{de}$ can cross the boundary condition $-1$, like quintom. From Fig. \ref{fa}(b) one can find that the coupling constant can affect the value of $w_{de}$ at high redshift greatly, but weaker at the low redshift. In Fig. \ref{fa}(c), we find that the proportion parameters of relativistic DM in the total DM can affect the values of $w_{de}$ in the whole evolution: the bigger $\gamma$, the smaller $w_{de}$ at high redshift and the bigger $w_{de}$ at low redshift.

The deceleration parameter, which is used to differentiate when the Universe transits from the decelerating phase to the accelerating phase, is given by $$q=-\frac{\ddot{a}}{aH^2}=-1-\frac{\dot{H}}{H^2}.$$
Taking $H=H_0h$ into account, $q$ can be rewritten in terms of $h^2$ as
\begin{eqnarray}
&q= -1+\frac{1+z}{2h^2} \frac{dh^2}{dz}.
\label{eq:q}
\end{eqnarray}
Using Eq. (\ref{eq:h2z}), the above equation can be written as
\begin{eqnarray}
&q=\frac{1}{4}\frac{(3m_1-4)c_1(1+z)^{\frac{3}{2}m_1}+ (3m_2-4)c_2(1+z)^{\frac{3}{2}m_2}}{c_1(1+z)^{\frac{3}{2}m_1}+c_2(1+z)^{\frac{3}{2}m_2} +(c_3-\Omega_{k0})(1+z)^2}.
\label{eq:qz}
\end{eqnarray}
\begin{figure*}
\centering
   \includegraphics[scale=.59]{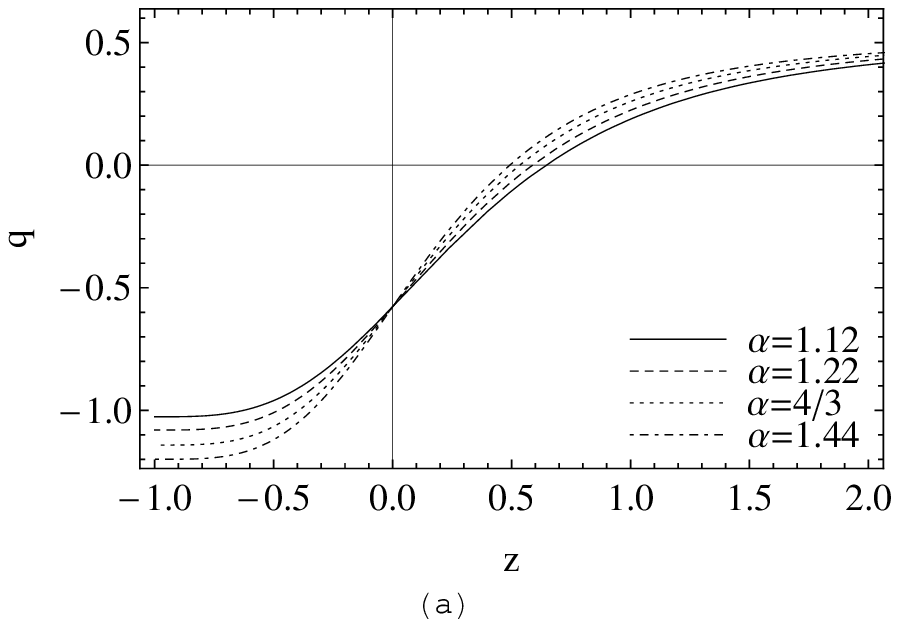}
   \includegraphics[scale=.59]{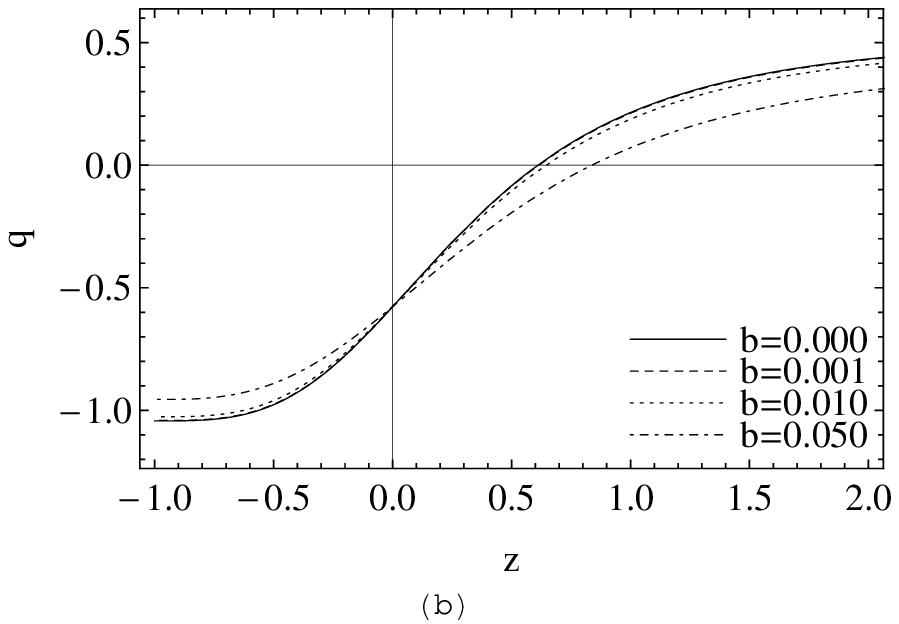}
   \includegraphics[scale=.59]{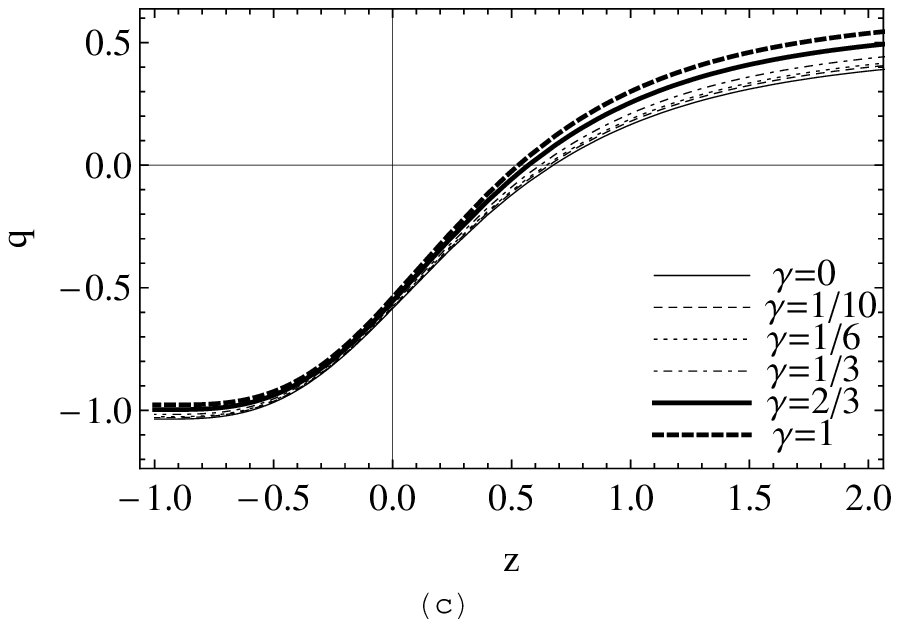}
\caption{The evolution of the deceleration parameter $q$ with respect to the redshift $z$ under different cases: (a) $b=0.01$, $\gamma=1/6$; (b) $\alpha=1.12$, $\gamma=1/6$; (c) $\alpha=1.12$, $b=0.01$. Here, we choose $\Omega_{k0}=0.02$, $w_{rdm}=0.1$, and $\Omega_{de}=0.73$.}
\label{fb}
\end{figure*}
Figure \ref{fb} shows the evolution of the deceleration parameter $q$ with respect to the redshift $z$. Figure \ref{fb}(a) is for variable $\alpha$ with the coupling constant $b=0.01$ and the proportion parameter $\gamma=1/6$, Fig. \ref{fb}(b) is for variable $b$ with $\alpha=1.12$ and $\gamma=1/6$, and Fig. \ref{fb} (c) is for different proportion parameters with $\alpha=1.12$ and $b=0.01$. From Fig. \ref{fb}, we find the deceleration parameter decreases from about $0.5$ to smaller than $-1$ as $z$ decreases in the whole evolution. The deceleration parameter is positive at high redshift, which indicates the earlier decelerating phase of the Universe. On the other hand, the negative deceleration parameter at low redshift indicates the accelerating phase of the Universe. Figure \ref{fb} shows that in the recent past at $z_{\text{T}}\simeq 1/2$, the Universe transits from the decelerating phase to the accelerating phase. Fig. \ref{fb} (a) shows that as $\alpha$ decreases, the transition occurs at relatively larger values of redshift, which indicates that the Universe enters the accelerating phase more early. Similarly, Fig. \ref{fb} (b) shows that as the coupling constant $b$ increases, the transition occurs earlier. And from Fig. \ref{fb} (c) one can find that as the proportion parameter of relativistic DM increases, the transition occurs later. With the observation data from SNe+CMB, the $\Lambda$CDM model gives the transition range $z_{\text{T}}=0.50\thicksim0.73$. However, with the help of Fig. \ref{fb} and our calculations, we find that the transition range is about $z_{\text{T}}=0.49 \thicksim 0.67$, which shows that the transition of the Universe from decelerating to accelerating expansion is close to that in the $\Lambda$CDM model. For this model, if the proportion parameter $\gamma$ is fixed, the present value of the deceleration parameter is a constant, which has nothing to do with the free parameters $\alpha$ or $b$. For $\gamma=1/6$, $q_0=-0.5775$. For fixed $\alpha$ and $b$, $q_0$ increases as $\gamma$ increases.

\begin{figure*}
\centering
   \includegraphics[scale=.59]{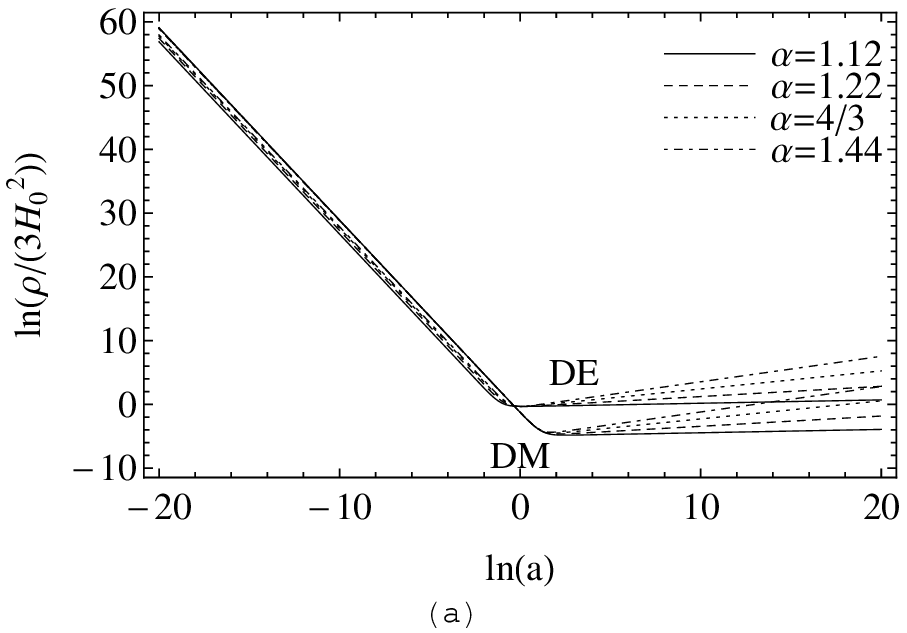}
   \includegraphics[scale=.59]{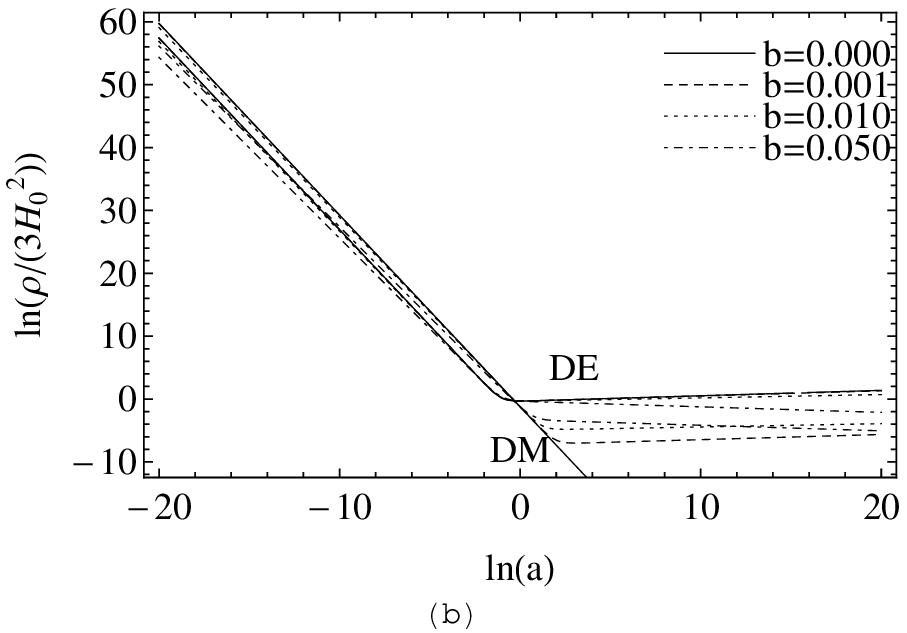}
   \includegraphics[scale=.59]{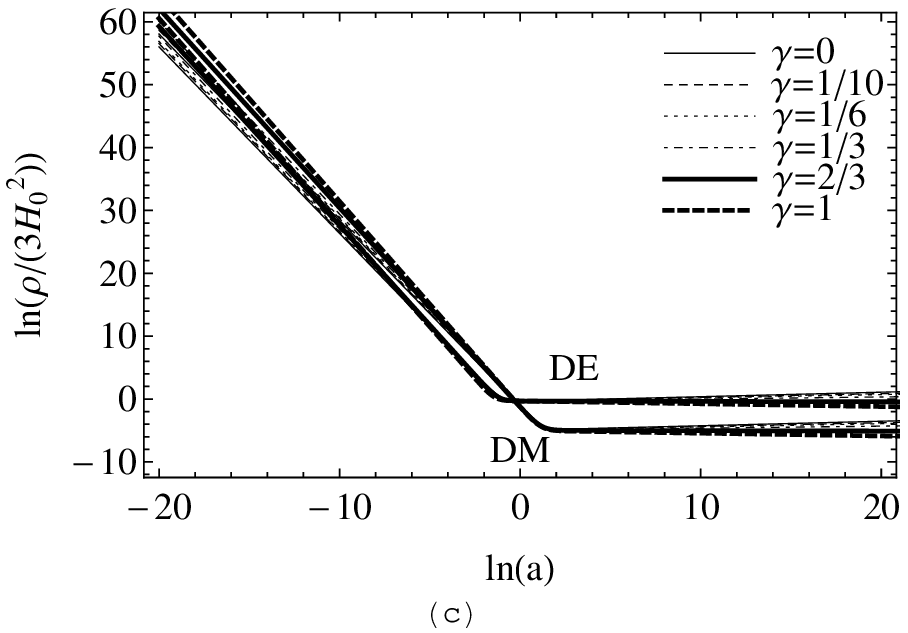}
\caption{The evolution of the energy density with respect to $\ln a$ under different cases: (a) $b=0.01$, $\gamma=1/6$; (b) $\alpha=1.12$, $\gamma=1/6$; (c) $\alpha=1.12$, $b=0.01$. Here, we choose $\Omega_{k0}=0.02$, $w_{rdm}=0.1$, and $\Omega_{de}=0.73$.}
\label{fc}
\end{figure*}
In order to examine how the densities of DM and MHRDE change over time, the evolution of the densities of DM and MHRDE are plotted in Fig. \ref{fc}. Combining the three graphics in Fig. \ref{fc}, it is easy to see that, in the past universe or at high redshift, the density of DM and MHRDE was comparable with each other; at low redshift, the MHRDE is dominating, which indicates that the accelerating expansion begins in the recent past, which is helpful in alleviating the coincidence problem.

\subsection{Statefinder diagnostic}
\label{ssf}

\begin{figure*}
\centering
   \includegraphics[scale=.58]{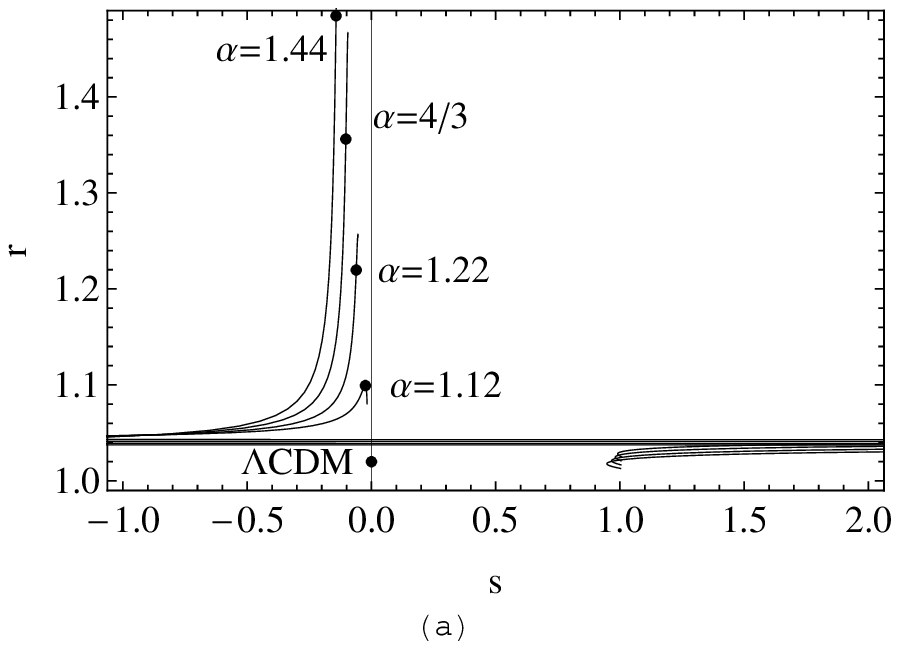}
   \includegraphics[scale=.58]{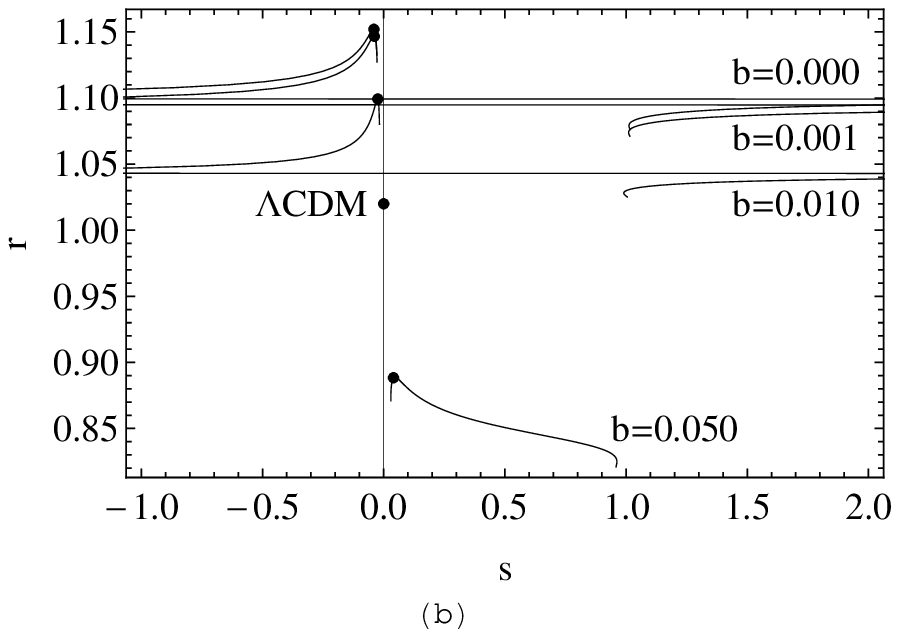}
   \includegraphics[scale=.58]{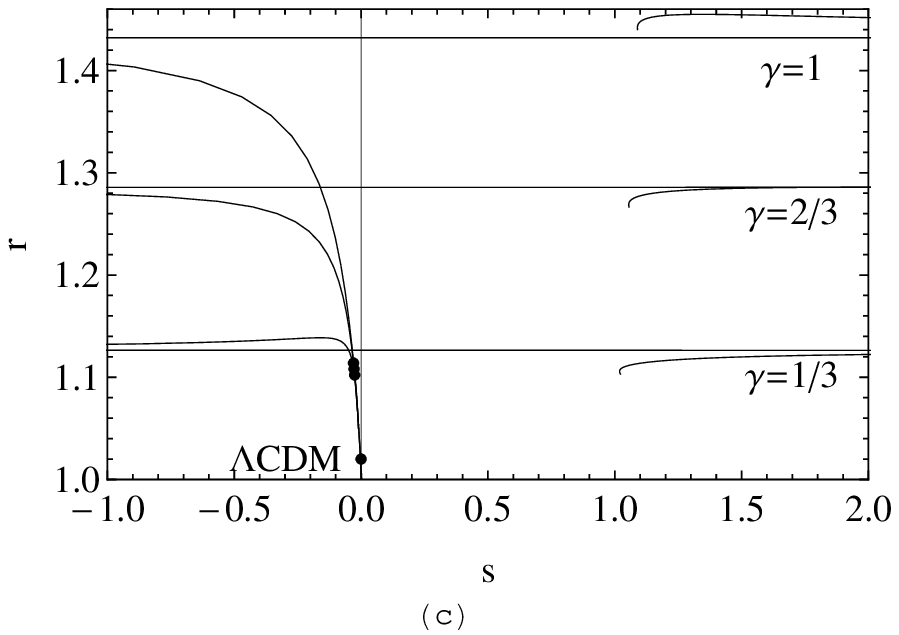}
   \includegraphics[scale=.58]{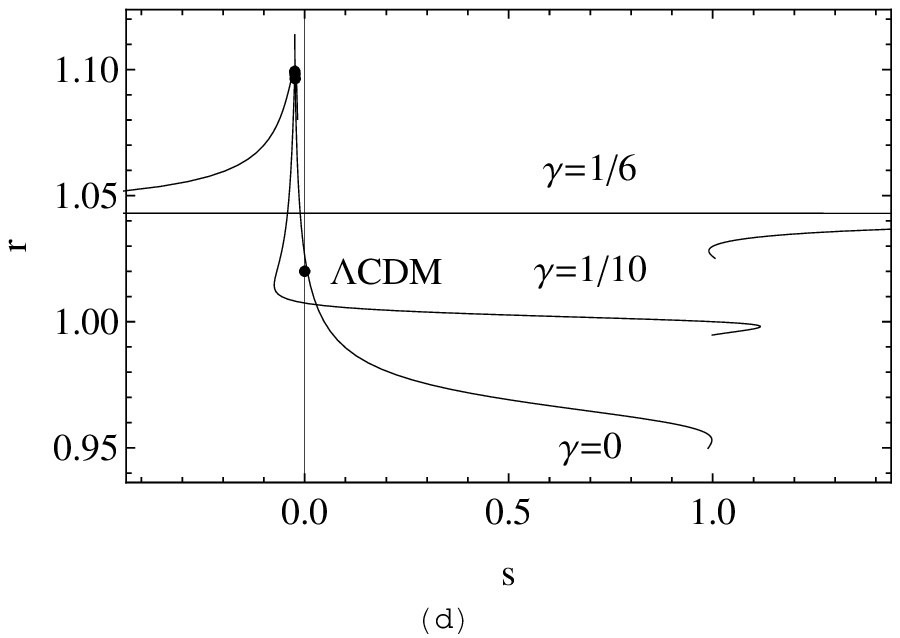}
   \includegraphics[scale=.58]{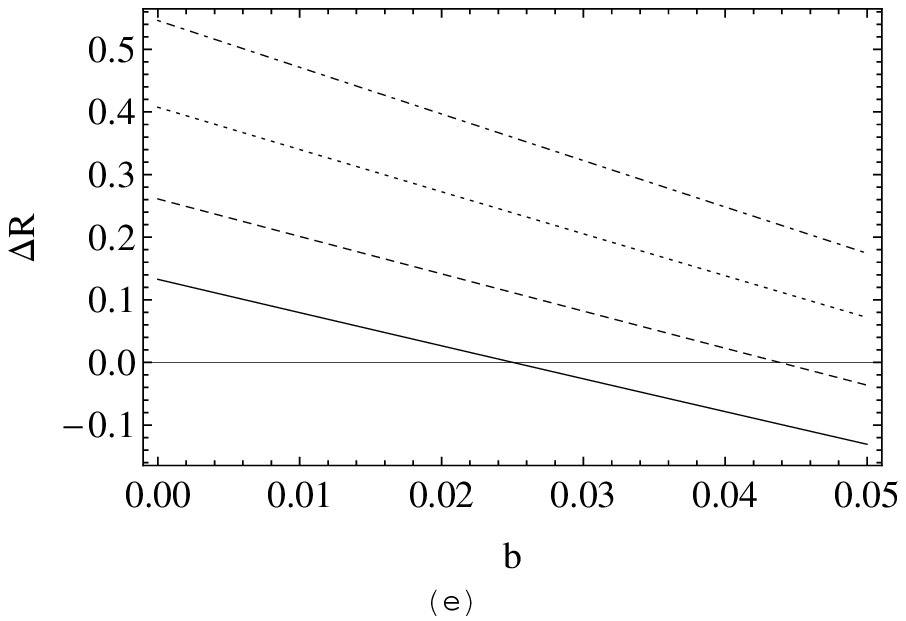}
   \includegraphics[scale=.58]{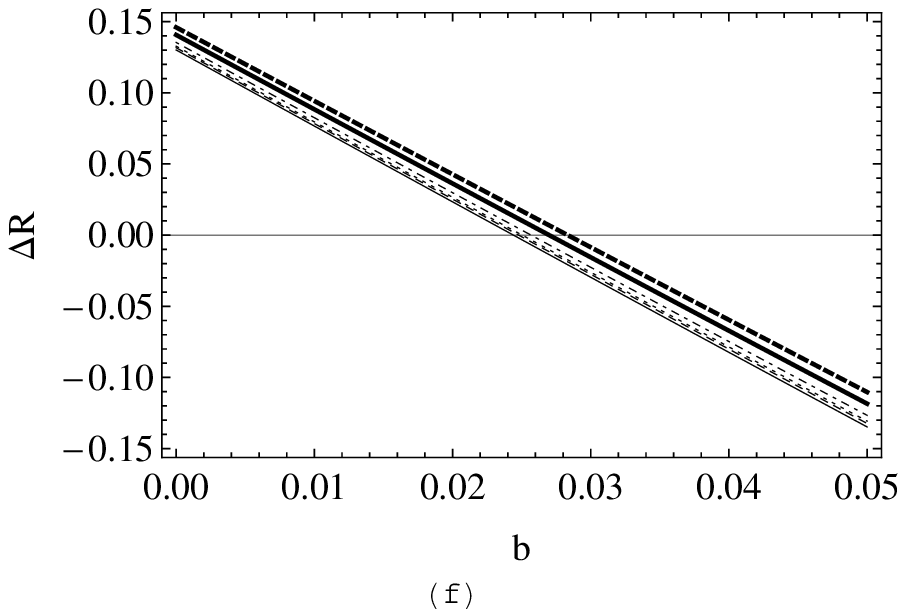}
\caption{Figures (a)-(d) are for the $r$-$s$ plane of IMHRDE model in different cases: (a). $b=0.01$ and $\gamma=1/6$, (b). $\alpha=1.12$ and $b=0.01$, (c) and (d) $\alpha=1.12$ and $b=0.01$. Figures (e) and (f) are today's distances between different IMHRDE model and $\Lambda$CDM model with respect to the coupling constant $b$, where $\Delta R =\sqrt{r_0^2+s_0^2}-1.02$. Figure (e) is for different $\alpha$ with $\gamma=1/6$, curves from below to top are for $\alpha=1.12, 1.22, 4/3, 1.44$. Figure (f) is for different $\gamma$ with $\alpha=1.12$, curves from top to below are for $\gamma=0, 1/10, 1/6, 1/3, 2/3, 1$. Here, we choose $\Omega_{k0}=0.02$, $w_{rdm}=0.1$ and $\Omega_{de0}=0.73$.}
\label{fd}
\end{figure*}

Since there are so many DE models that have been put forward, in order to discriminate different DE models from each other, Sahni $\it{et}$ $\it{al.}$ \cite{Sahni2003} introduced the statefinder pair $\{r,s\}$, which uses the third time derivative of the scale factor $a(t)$, to diagnose and discriminate behaviors of different DE models. In this section we would like to check up on the statefinder pair in the IMHRDE model. The statefinder pair $\{r,s\}$ is given by \cite{Sahni2003,Evans2005,Setare2007}
\begin{eqnarray}
&r=\frac{\dddot{a}}{aH^3},\quad s=\frac{r-\Omega_{tot}}{3(q-\Omega_{tot}/2)},\label{eq:rs}
\end{eqnarray}
where $q$ is the deceleration parameter and $\Omega_{tot}=\Omega_{dm}+\Omega_{de}=1+\Omega_{k0}(1+z)^2/h^2$. From Eqs. (\ref{eq:q}) and (\ref{eq:rs}), the statefinder pair can be written in the following form
\begin{eqnarray}
&r=1 +\frac{3} {2h^2}\frac{dh^2}{dx} +\frac{1}{2h^2} \frac{d^2h^2}{dx^2},\\
&s=- \frac{1}{3} \frac{ (h^2)'' +3(h^2)' -2\Omega_{k0} e^{-2x}}{ (h^2)' +3h^2 +\Omega_{k0}e^{-2x}}.
\label{eq:r}
\end{eqnarray}
Thus, using Eq. (\ref{eq:h2z}), we obtain
\begin{eqnarray}
  &r=\frac{1}{2} \frac{\eta_1}{\eta_2},\quad s=\frac{1}{3} \frac{\eta_3}{\eta_4},
\end{eqnarray}
where
\begin{eqnarray}
  &\eta_1= (2- \frac{9}{2} m_1+ \frac{9}{4} m_1^2)c_1 (1+z)^{\frac{3}{2} m_1}
    +(2- \frac{9}{2} m_2+ \frac{9}{4} m_2^2)c_1 (1+z)^{\frac{3}{2} m_2},\\
  &\eta_2= c_1 (1+z)^{\frac{3}{2} m_1} +c_2 (1+z)^{\frac{3}{2} m_2}
    + (c_3-\Omega_{k0}) (1+z)^2,\\
  &\eta_3=  9m_1(m_1-2) c_1 (1+z)^{\frac{3}{2}m_1 }
     +9m_2(m_2-2) c_2 (1+z)^{\frac{3}{2}m_2 }
   -8c_3 (1+z)^2,\\
  &\eta_4= 6(m_1-2) c_1 (1+z)^{\frac{3}{2}m_1 }
    +6m_2(m_2-2) c_2 (1+z)^{\frac{3}{2}m_2 }
    -4c_3 (1+z)^2.
\end{eqnarray}
\begin{figure}
\centering
   \includegraphics[scale=.5]{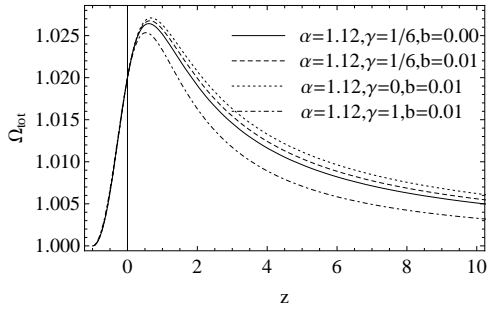}
\caption{The curves of $\Omega_{tot}$ with respect to $z$ under different cases. Here, we choose $\Omega_{k0} =0.02$, $w_{rdm} =0.1$ and $\Omega_{de0} =0.73$.}
\label{fg}
\end{figure}
The $r$-$s$ evolutionary trajectory in the IMHRDE model is shown in Fig. \ref{fd}. In the statefinder $r-s$ plane, the $\Lambda$CDM model in the flat universe corresponds to a fixed point $\{r,s\}=\{1,0\}$, and the other models of DE's behaviors can be measured from the distance between them and the $\Lambda$CDM point. But now, in the nonflat universe, the $\{r,s\}$ pair would be $\{\Omega_{tot}, 0\}$, where $\Omega_{tot} =1+ \frac{\Omega_{k0}(1+z)^2}{h^2}$ varying with time, then the statefinder pair would not be a fixed point but a line segment. Today's state of the evolution of the $\Lambda$CDM model is $\{r_0,s_0\} =\{1.02,0\}$. In Fig. \ref{fg}, we have plotted the total energy density $\Omega_{tot}$ with respect to the redshift. Figure \ref{fg} shows that different parameters could strongly influence the trend of the curves of $\Omega_{tot}$ at high redshift but much weaker at low redshift.

The black spots in Fig. \ref{fd} correspond to today's state for the $\Lambda$CDM model and the IMHRDE model. From Fig. \ref{fd}(a), we find that as the value of $\alpha$ becomes bigger, the range of the trajectory becomes larger. Figure \ref{fd}(b) shows that the present values of statefinder pair under different coupling constants are $\{r_0, s_0\} =$ $\{1.152, -0.0404\}$, $\{1.147, -0.0388\}$, $\{1.099, -0.0243\}$, $\{0.888, 0.0403\}$, respectively. It is easy to find that as $b$ increases, the range of the trajectory becomes smaller first and then bigger. Figures \ref{fd}(c) and \ref{fd}(d) show that the influence of $\gamma$ on the range of the trajectory is weak. Today's distances between different IMHRDE models and $\Lambda$CDM models with respect to the coupling constant $b$ are plotted in Figs. \ref{fd}(e) and \ref{fd}(f), where $\Delta R =\sqrt{r_0^2+s_0^2}-1.02$. From these two graphics, one can find that as $b$ increases the distance becomes smaller first and then bigger.

Figure \ref{fd}(a) shows that, when $s\rightarrow\pm\infty$, $r\rightarrow r_{const}$ ($r_{const}$ is a positive constant). In Fig. \ref{fd}(a), it is easy to see that, as $s$ increases from $-\infty$ to $0$, $r$ increases from $r_{const}$ to a finite value, and as $s$ increases from $0$ to $+\infty$, $r$ increases from a positive value less than $r_{const}$ to $r_{const}$. Figure \ref{fd}(b) shows that, for $b=0, 0.001, 0.01$, as $s$ increases from $-\infty$ to $+\infty$, $r$ increases from $r_{const}$ to a maximum first, and then decreases from the maximum to a constant bigger than $r_{const}$ when $s<0$, but for $s>0$, $r$ increases from a small positive value to $r_{const}$ as $s$ increases. When $b=0.005$, as $s$ increases $r$ increases first and then decreases. Figures \ref{fd}(c) and \ref{fd}(d) show that for big values of the proportion parameters, one can find that as $s$ increases from $-\infty$ to $0$, $r$ decreases from $r_{const}$ to a finite value, and when $s$ is big enough, as $s$ increases, $r$ increases from a positive value to $r_{const}$. However, if there is no relativistic DM or its proportion is small, the phenomena $s\rightarrow \pm\infty, r\rightarrow r_{const}$ cannot happen.

\begin{figure*}
\centering
   \includegraphics[scale=.58]{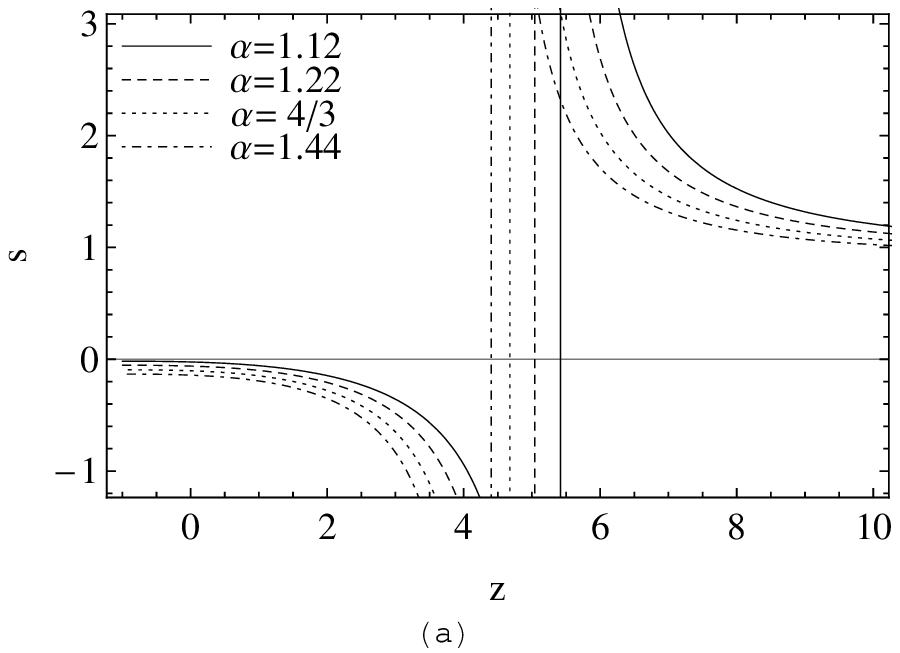}
   \includegraphics[scale=.58]{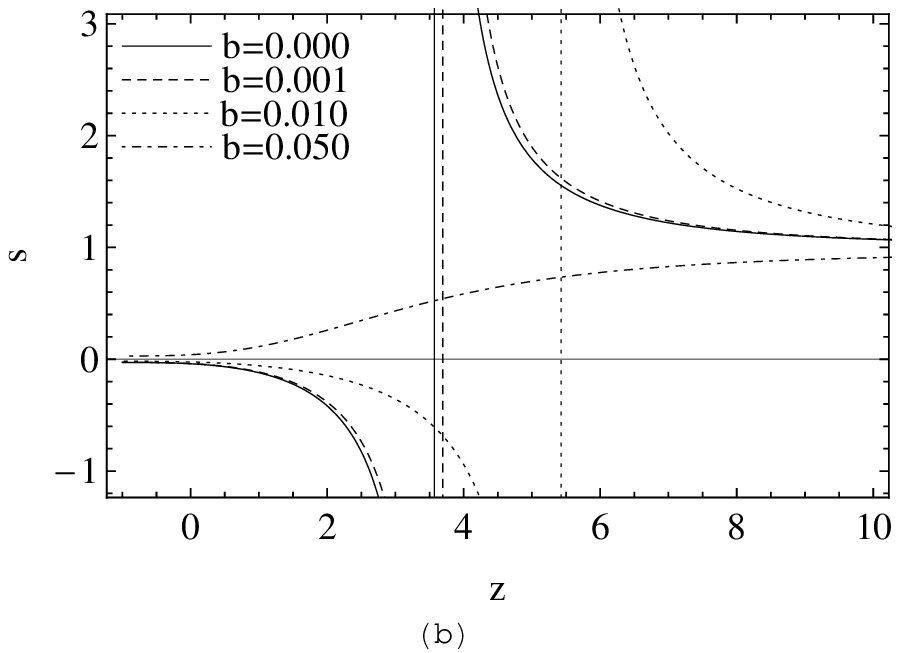}
   \includegraphics[scale=.58]{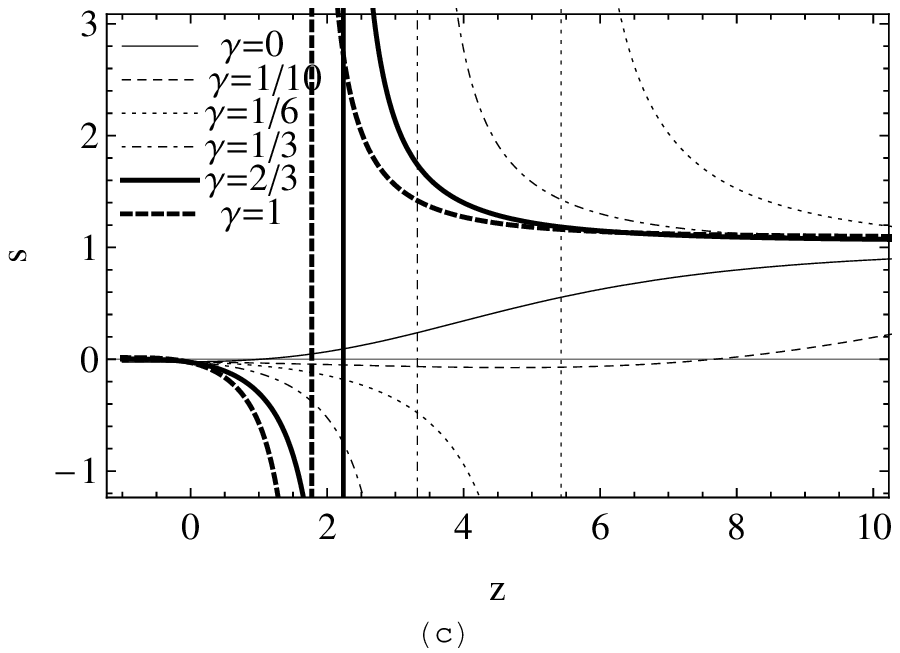}
   \includegraphics[scale=.58]{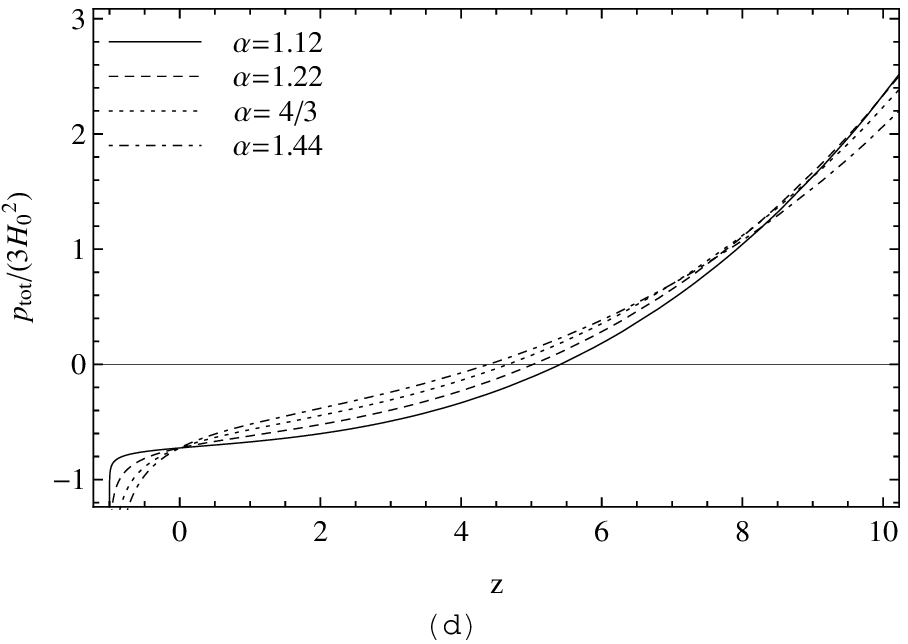}
   \includegraphics[scale=.58]{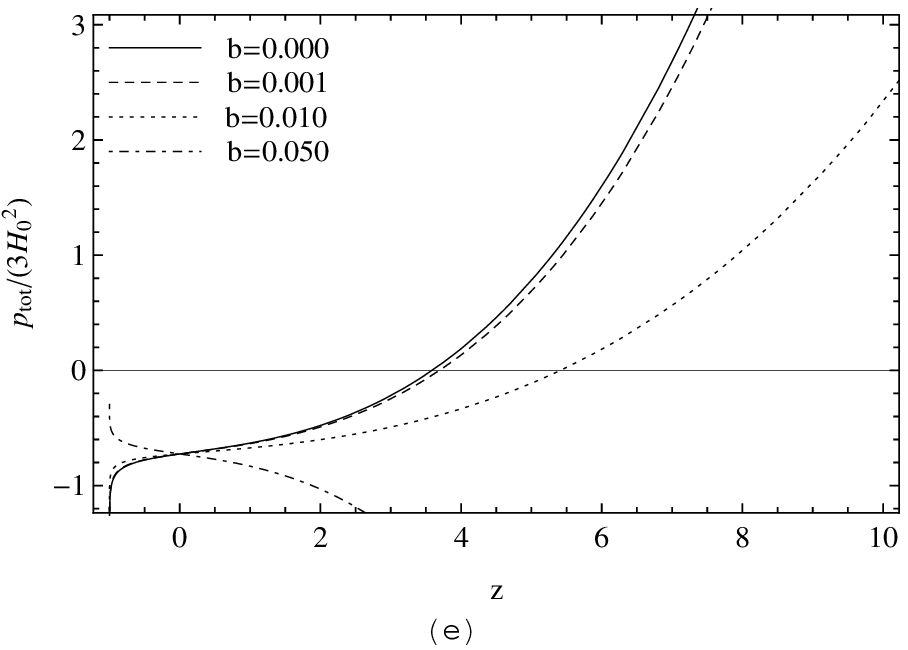}
   \includegraphics[scale=.58]{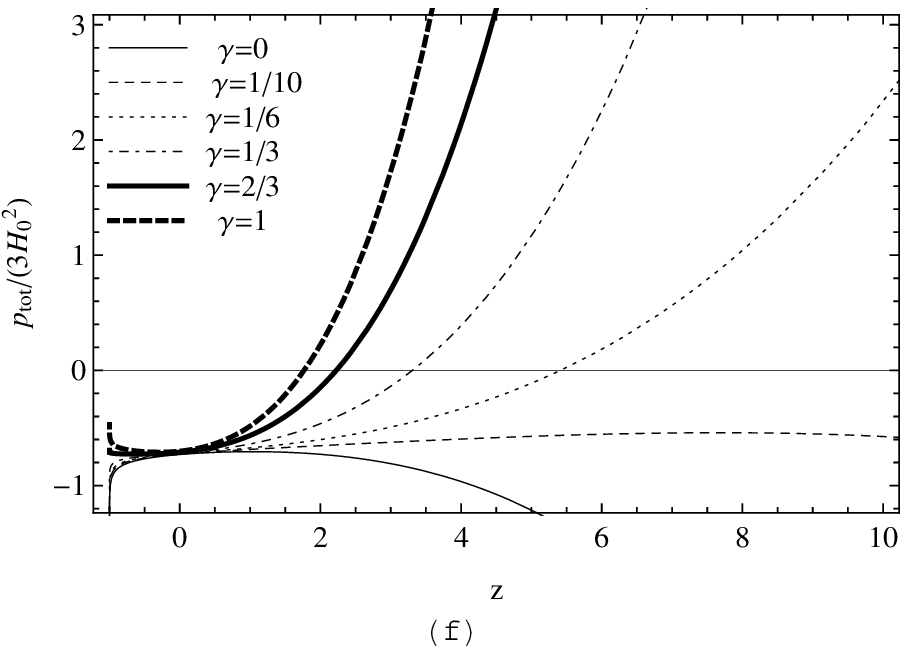}
\caption{Figures (a), (b), and (c) are for the evolution of the statefinder parameter $s$ with respect to $z$ under different cases: (a) $b=0.01$, $\gamma=1/6$; (b) $\alpha=1.12$, $\gamma=1/6$; (c) $\alpha=1.12$, $b=0.01$. Figures (d), (e) and (f) are for the evolution of the total pressure $p_{tot}$ with respect to $z$ under different cases: (d) $b=0.01$, $\gamma=1/6$; (e) $\alpha=1.12$, $\gamma=1/6$; (f) $\alpha=1.12$, $b=0.01$. Here, we choose $\Omega_{k0}=0.02$, $w_{rdm}=0.1$ and $\Omega_{de0}=0.73$.}
\label{fe}
\end{figure*}

In order to explain the above phenomena, let us take the total density $\rho_{tot}$ and total pressure $p_{tot}$ into consideration; then the statefinder pair can be written in the following form:
\begin{eqnarray}
  &r=&\Omega_{tot}+\frac{9}{2} \frac{p_{tot}+\rho_{tot}}{\rho_{tot} -3 \frac{k}{a^2}} \frac{\dot{p}_{tot}}{\dot{\rho}_{tot}},\\
  &s=&\frac{p_{tot}+\rho_{tot}}{p_{tot}} \frac{\dot{p}_{tot}}{\dot{\rho}_{tot}}.
  \label{eq:rsp}
\end{eqnarray}
From Eq. (\ref{eq:rsp}) one can find that the statefinder $s$ is exceedingly sensitive to the total pressure $p_{tot}$. At a very early time, which is the dark  matter dominating stage, the positive pressure of the relativistic DM ensures that the total pressure in the Universe is positive, which is just like radiation, whose pressure is positive, then we call this stage the ``radiation stage''. Much later, the Universe would be DE dominated, whose negative pressure can drive the Universe to accelerating expansion, for this stage we call it the ``accelerating expansion stage''. Between these two stages, one can find that there would be a precise moment or a stage in which the positive pressure of relativistic DM is balanced by the negative pressure of DE. At such a moment or stage, one has $p_{tot} \simeq 0$, $|s| \rightarrow \infty$; in this paper, we call such stage the ``dust stage'' \cite{Alam2003,Debnath2013}. From Fig. \ref{fe}, one can find that when $s\rightarrow \pm\infty$, $p_{tot}=0$.

According to Figs. \ref{fe}(a) and \ref{fe}(d), one can say that as $\alpha$ decreases, the dust stage occurs early. Figures \ref{fe}(b) and \ref{fe}(e) show that if the coupling constant is too big, the dust stage disappears, and as the value of $b$ increases the dust stage occurs earlier. In Fig. \ref{fe}(c) and Fig. \ref{fe}(f) we choose $\alpha=1.12$ and $b=0.01$. In this case, we find that as the relativistic DM's proportion decreases the dust stage appears earlier. However, if the proportion is too small, the dust stage cannot appear. Then we can say that the existence of the dust stage in the past and the time of its occurrence are all influenced not only by $\alpha$ and $\gamma$, but also by the coupling constant $b$. Above all, we can say that if the parameters are suitable, the IMHRDE model could explain the Universe's transition from radiation stage to accelerating expansion stage through the dust stage.

\begin{figure}
\centering
   \includegraphics[scale=.48]{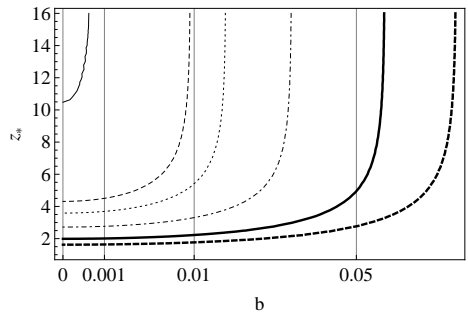}
\caption{Redshift for the Universe entered the dust stage with respect to $b$ in the non-uniform coordinate, curves from left to right is for $\gamma=0, 1/10, 1/6, 1/3, 2/3, 1$ and $\alpha=1.12$. Here, we choose $\Omega_{k0}=0.02$, $w_{rdm}=0.1$ and $\Omega_{de0}=0.73$.}
\label{fh}
\end{figure}
Now, we turn our attention to the precise moment when the dust stage occurs. Figure \ref{fh} shows the precise moment $z_*$ with respect to $b$ for different $\gamma$. From Fig. \ref{fh}, it is easy to find that the coupling constant $b$ and proportion parameter $\gamma$ all have great influence on $z_*$. For a fixed $\gamma$, as $b$ increases, $z_*$ increases more and more rapidly and tends to infinity finally. One can also find that, for a fixed $\gamma$, only when the coupling constant is small enough can the dust stage appear. Here we give some values of $z_*$: for $\alpha=1.12$, $\gamma=1/10$ and $b=0.001$, we have $z_*=4.505$; for $\alpha=1.12$, $\gamma=1/10$ and $b=0.005$, we have $z_*=5.688$; for $\alpha=1.12$, $\gamma=1/10$ and $b=0.009$, we have $z_*=10.99$.

\subsection{$Om$ diagnostic}

Now, in this section, we will turn to the $Om$ diagnostic, which is helpful in distinguishing different DE models without referencing to either the matter density or $H_0$ \cite{Sahni2008}. It is defined as
\begin{eqnarray}
&Om(y)=\frac{h^2(y)-1}{y^3-1},
\label{eq:Om}
\end{eqnarray}
where $y=1+z$ and $h(y)=H(y)/H_0$. From the definition, we see that $Om$ involves only the first derivative of the scale factor through the Hubble parameter. Using Eq. (\ref{eq:h2z}) we obtain
\begin{eqnarray}
&Om(z)=\frac{c_1(1+z)^{\frac{3}{2}m_1}+c_1(1+z)^{\frac{3}{2}m_1}+(c_3-\Omega_{k0})(1+z)^2-1}{(1+z)^3-1}.
\label{eq:Omz}
\end{eqnarray}
\begin{figure*}
\centering
   \includegraphics[scale=.59]{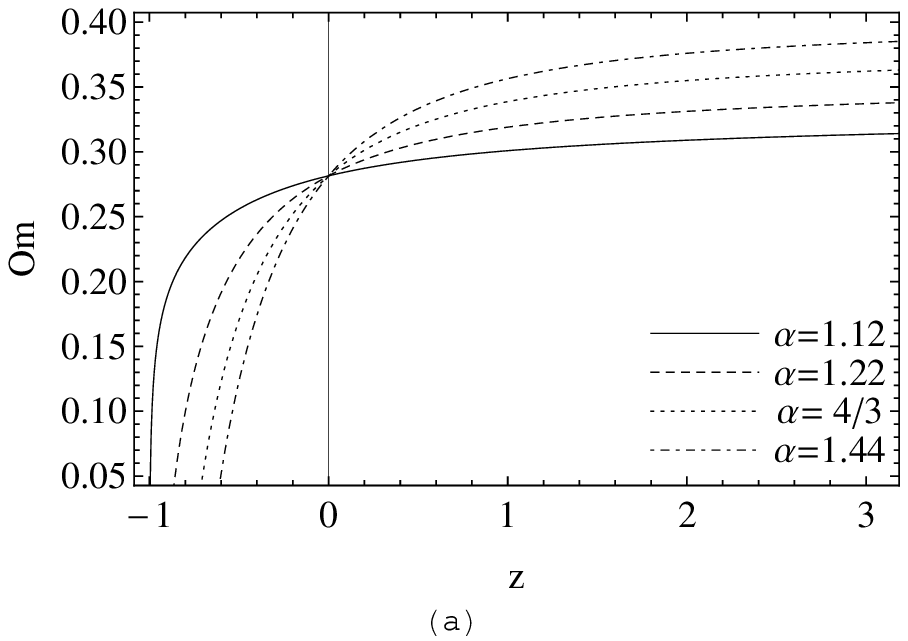}
   \includegraphics[scale=.59]{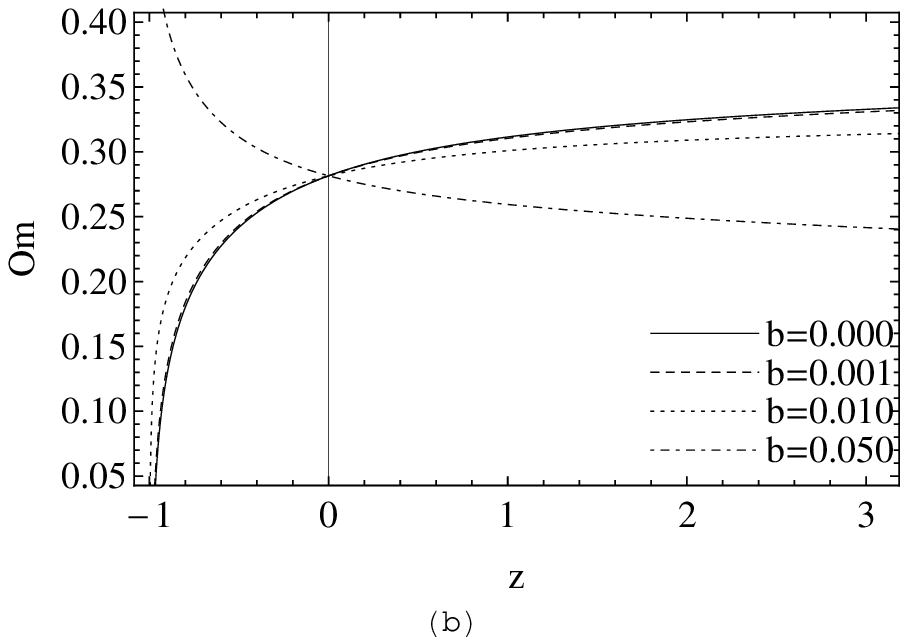}
   \includegraphics[scale=.59]{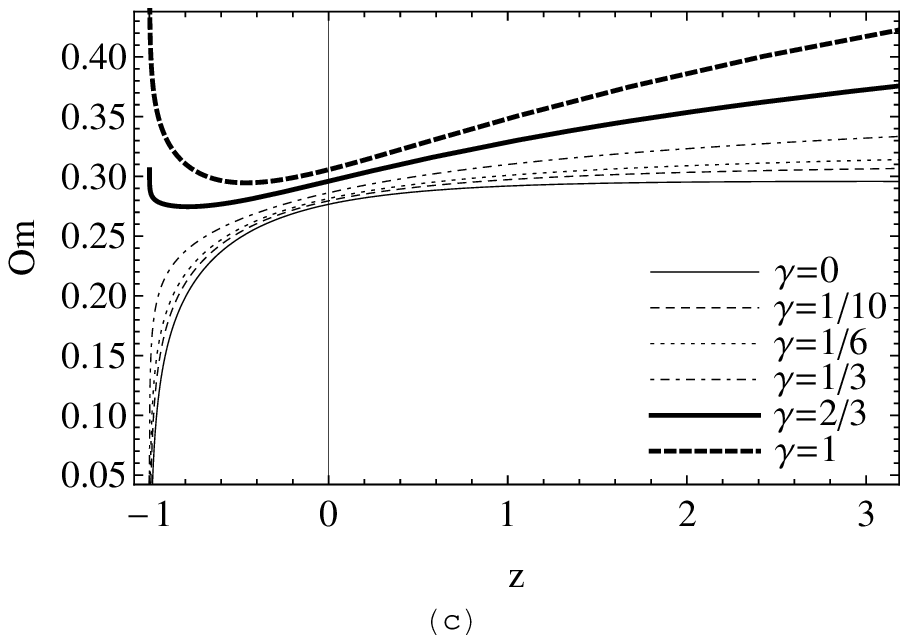}
\caption{The curves of $Om(z)$ with respect to $z$ under different cases: (a) $b=0.01$, $\gamma=1/6$; (b) $\alpha=1.12$, $\gamma=1/6$; (c) $\alpha=1.12$, $b=0.01$. Here, we choose $\Omega_{k0}=0.02$, $w_{rdm}=0.1$ and $\Omega_{de0}=0.73$.}
\label{ff}
\end{figure*}
Figure \ref{ff} shows the variations of $Om(z)$ with  respect to the redshift $z$. According to the work \cite{Sahni2008}, an upward trend of $Om(z)$ represents phantom $(w_{de}<-1)$ and a downward trend of $Om(z)$ represents quintessence $(w_{de}>-1)$. In Fig. \ref{ff}(a) we find that for $\alpha=$ $1.12$, $1.22$, $4/3$, and $1/44$, $Om(z)$ increases as the redshift $z$ increases, which is similar to the phantom model. Figure \ref{ff}(b) shows that when $b =0$, $0.001$ and $0.01$, the $Om(z)$ curves have an upward trend, but it is a downward trend for $b=0.05$. Thus, we can say that for $\alpha=1.12$ and $\gamma=1/6$, the IMHRDE acts like quintessence when the interaction is very weak, but it is phantom-like under strong interaction. From Fig. \ref{ff}(c), we find for $\gamma=2/3$ and $\gamma=1$, as $z$ increases $Om(z)$ decreases first, and then increases, which indicates that the state of the IMHRDE model will transit from quintessence to phantom phase. When the proportion parameter $\gamma$ is small, $Om(z)$ has a positive slope only, suggesting the IMHRDE model is phantom-like.

\section{Conclusion}
\label{scon}

In this paper, we have examined the evolution of a nonflat FRW universe, which is filled with DM and MHRDE. The IR cutoff of the MHRDE is given by the modified Ricci scalar; its form is $\rho_{de}= \frac{2}{\alpha-\beta} \left( \dot{H} +\frac{3}{2} \alpha H^2 \right)$. The total DM has two components: relativistic DM and non-relativistic DM; the EOS parameters of the relativistic DM and non-relativistic DM are chosen as $w_{rdm}= 0.1$ and $w_{nrdm} =0$, respectively. The present density ratio between DM and DE and the present EOS parameter of DE are chosen as the boundary conditions; then the parameter $\beta$ can be obtained in terms of the free parameter $\alpha$. In the present paper, we take the form of the interaction between DM and MHRDE as $Q=3bH(\rho_{de}+\rho_{dm})$.

In order to examine the evolution of the nonflat universe, the EOS parameter of MHRDE and the deceleration parameter are studied  first. We find that the value of $\alpha$ have a great effect on the future value of $w_{de}$, $b$ has a great effect on the past value of $w_{de}$ and the proportion parameter $\gamma$ could influence the value of $w_{de}$ in the past and the future. Figure \ref{fa} also shows that the MHRDE behaves like DM in the early universe and phantom-like in the future. By examining the deceleration parameter, we find that the Universe's transition from decelerating to accelerating expansion would be affected by the values of $\alpha$, $b$, and $\gamma$. We also find that the transition of the Universe from decelerating to accelerating expansion is close to that in the $\Lambda$CDM model. Combining the evolution of the densities of DM and MHRDE, we find that MHRDE's density is comparable to DM's at high redshift and MHRDE is dominating at low redshift, which indicates that the accelerating expansion begins in the recent past. This is helpful in solving the coincidence problem.

Next, we have studied the statefinder diagnostic for the IMHRDE model by plotting the trajectories in the $r-s$ plane. One can find that for some parameters we choose, a special phenomena appears: for $s\rightarrow \pm \infty$, $r\rightarrow r_{const}$. In order to clarify the phenomena, we take the total density and the total pressure into consideration. As we know, the statefinder parameter $s$ is sensitive to the total pressure $p_{tot}$; then we studied the evolution of the total pressure $p_{tot}$ next. One can find that for $s\rightarrow \pm\infty$, $p_{tot} =0$, and we call this stage the ``dust stage''. Whether there is a dust stage is decided by $b$ and $\gamma$ and the moment when the dust stage occurs due to $\alpha$, $\beta$ and $\gamma$. For suitable model parameters, one can say that the IMHRDE model can explain the Universe's transition from the radiation stage to the accelerating expansion stage through the dust stage. By studying the $Om$ diagnostic we find that if the interaction between DM and MHRDE is weak and the proportion of relativistic DM in the total DM is small, the IMHRDE would be phantom-like. Strong interaction can lead to quintessence-like and large $\gamma$ can lead to the transition from quintessence to phantom.

\section*{Acknowledgments}

This work was supported in part by the Talent  Cultivation Foundation of Kunming University of Science and Technology under Grant No. KKSY201207053. Y. Zhang would like to acknowledge the support of the working Funds of the Introduced High-level Talents of Yunnan Province from the Department of Human Resources and Social Security of Yunnan Province.

\end{document}